\documentclass[12pt]{article}
\usepackage{curves}
 \begin{document}
\thispagestyle{empty}
 \rightline{EFI 97-14}

 \begin{center} \large {\bf T-Duality Considerations for Hadronic Strings}

 \medskip\normalsize 
James A. Feigenbaum, Peter G.O. Freund and Mircea Pigli\\

             {\em Enrico Fermi Institute and Department of Physics\\
             The University of Chicago, Chicago, IL 60637, USA}

 \end{center}
 \medskip

\noindent{\bf Abstract}: 
We show that a simple account of hadronic high energy total cross-sections 
and of the observed 
masses and flavor-gauge-like couplings of the familiar vector mesons is 
obtained in a hadronic string model suggested by QCD in the limit of a large
number of colors. Our picture involves a Minkowski space which in addition
to the well established continuous one time and three space dimensions 
is endowed with an extra {\em discrete} space dimension involving a 
minimal 2-point lattice with spacing of order $10^{-14}$ cm in one of the two 
T-dual pictures. New mesonic states with characteristic decay modes are 
expected in such a picture.
\setcounter{page}{0}
\newpage

\bigskip

{\bf 1. Introduction}

\bigskip

With the recent understanding of the role of dualities
in string theory \cite{RD, DR}, 
the hitherto less studied open strings with Chan-Paton (CP)
rules \cite{GSW}
have regained interest. These CP rules were originally abstracted from 
hadron phenomenology, where they form part of the understanding of high
energy hadronic total cross-sections. In QCD, the theory of strong 
interactions, hadrons can be pictured as tubes of color-flux capped
by quarks (or antiquarks), or in other words as open strings with quarks at 
their ends.
We wish to explore here whether the recent advances in the theory of open 
strings may shed some light on the physics of hadrons. We shall see that some 
interesting new perspectives on some old hadron problems can be
gained this way, and that an unexpected 
connection between quark flavors and the structure of Minkowski space-time 
seems to emerge.
In particular, we will find that the validity of hadronic Chan-Paton rules, 
(and of their successful consequences for high energy hadron scattering)
and the mass spectrum and flavor-gauge-like coupling pattern of vector 
mesons are readily explained. We will extend Minkowski space by 
adding to it a {\em discrete} space-like dimension involving a 
minimal 2-point lattice with spacing of order $10^{-14}$ cm in one of the two 
T-dual pictures. New mesonic states with characteristic decay modes are 
expected in such a picture.

\bigskip

{\bf 2. Open Strings}

\bigskip

Let us start from the CP rules, for which there is ample evidence coming 
primarily from high energy hadron scattering processes at fixed momentum
transfer. The Lie group involved in the hadronic CP rules is $U(N)$, $N$ 
being the 
number  of quark flavors ({\em not} colors). Consider a process for which, 
say both the $st$ and $su$
quark diagrams --- i.e. both quark diagrams involving the ``exotic" $s$ channel
--- are forbidden because the corresponding CP rule traces vanish \cite{PF,
HH, JR, HHR}. Examples of
such processes are $K^+K^+$-, $K^+p$-, $pp$-, $\pi^+ \pi^+$-, $\phi p$- 
scattering. 
At string tree level in all
these cases the contribution of the mesonic Regge poles to the scattering 
amplitude at large $s$ and fixed $t$ (in partcular $t=0$) is purely
real, so that the characteristic $s^{-\frac{1}{2}}$ contribution to the 
corresponding total cross-section is absent, in agreement with 
experiments (for an up to date analysis of the phenomenology see ref. 
\cite{FFP}).
The way this is achieved is by the degeneracy of the odd signature Regge
trajectory on which the $\rho$ and $\omega$ vector mesons lie with the 
even signature trajectory on which the tensor mesons $f$ and $a2$ lie.
This degeneracy is also experimentally confirmed and there are further 
instances in which the CP rules are confirmed.

These rules follow from QCD in the limit of a large number of colors in which
hadrons appear as strings
with quarks at their ends. For the purposes of this work we 
will assume this string picture to hold. Essentially this amounts to assuming
that the large number of colors approach is already reasonably good for three 
colors. There is independent evidence \cite{FR} in favor of such a string 
picture. Physically one can think of the strings as tubes of color flux
capped by quarks and antiquarks, as was mentioned above. These hadronic 
strings differ from the fundamental superstrings in that they have massive
quarks at their ends, in that they are not in a critical dimension and in
that they are not supersymmetric. In spite of all these differences,
we will apply T-duality to hadronic strings.

Consider a mesonic string $[ij]$ (baryons can be treated similarly) 
as an open string with a quark of flavor $i$ at one end and an antiquark of
flavor $j$ at the other end,
moving in 4-dimensional Minkowski space. Let us, for the moment, assume 
the existence of an ``extended" Minkowski space, which
in addition to its time and 3 noncompact space 
dimensions also has a fifth {\em continuous} space 
dimension compactified to a circle of radius $R$.
The appearance of such an extra space dimension may be connected with the 
number of colors going to infinity limit. But
a continuous compact fifth dimension
implies the existence of a tower of Kaluza-Klein (KK) states and this causes
serious problems especially in the
closed string glueball sector, which is flooded by KK states at all energies.
As we will see in section 4, this problem can be solved by extending 
Minkowski space by a
{\em discrete} rather than continuous fifth dimension. In this more 
modest extension the theory remains 4-dimensional, as expected for a finite
number of colors.
For now let us temporarily allow the introduction of a compact 
{\em continuous} fifth dimension and 
explore the consequences of this assumption for open strings.
This allows us to present in Section 2 
the open string phenomenology in a familiar setting, 
which does not get radically changed upon discretizing the fifth dimension.
By contrast, in Section 4 we explain the untenable situation which then 
develops in the closed string glueball sector and in Section 5 we deal
with this problem by switching to a discrete fifth dimension.

We now proceed to the discussion of open strings in the presence of a fifth 
dimension in the form of a circle of radius $R$.

Were we to ignore the quark masses, we could now switch to the T-dual picture
in which the radius of the compact dimension's circle is 
$$
  R'=\frac{\alpha '}{R},
\eqno(1)
$$ 
where $\alpha ' \approx 1$  GeV$^{-2}$ is the 
hadronic 
Regge slope. Equivalently, the tension of the hadronic string is 
$T= \frac{1}{2 \pi \alpha'}$. As far as the compact fifth dimension is 
concerned, the T-duals of the Neumann boundary conditions 
at the end of the original string are then Dirichlet boundary conditions for 
its dual. This dual string must end \cite{P} on one of a set of $N$ Dirichlet
3-branes, $N$ 
being the number of quark flavors. These $N D$-branes are $N$ copies of ordinary
4-dimensional Minkowski space, located at $N$ given values 
$\theta_1 R', \theta_2 R', ..., \theta_N R'$ of the compact fifth 
coordinate $\theta R'$.
The specific and not necessarily all distinct values of the $\theta_i$ 
reflect the details of the Wilson line included when compactifying the fifth 
coordinate. If all distinct, they lead to a breaking of the flavor $U(N)$ 
symmetry  to 
$U(1)^N$. The ``nondiagonal" {\em flavor} gauge bosons present as states of 
the string thus acquire masses in a pattern characteristic of an adjoint
representation Brout-Englert-Higgs (BEH) 
phenomenon. Specifically the mass $m_{ij}$ 
of the $[ij]$ vector meson is given by

$$ 
   m_{ij}^2= \left( \frac{(\theta_i - \theta_j) R'}{2\pi \alpha'} \right)^2
\eqno(2)
$$

 Thus the $K^*, \rho^{\pm}, D^*,
D^*_s, B^*$, etc... mesons acquire masses, whereas the\\
$\rho^0, \omega, \phi, J/\psi,
\Upsilon$ and toponium remain massless. This is obviously unacceptable and 
quite  reminiscent of the old Sakurai theory of flavor gauging \cite{S}. 
But in all 
of this we have ignored the masses of the quarks, acquired through the BEH
phenomenon of the Standard Model. When all the quark masses are set to zero
and all the $\theta_i$ are equal, then all $N^2$ vector mesons would be 
massless gauge bosons. But with {\em massive constituent} quarks
at the ends of the strings 
there is another contribution to $m_{ij}$, namely the sum $m_i+m_j$ of the 
masses  of the constituent quarks of flavor $i$ and $j$. Then

$$
  m_{ij}^2= (m_i+m_j)^2 +  \left( \frac{(\theta_i - \theta_j) R'}{2\pi \alpha'} \right)^2.
\eqno(3)
$$

This is a much more interesting and realistic formula. The first term on the 
right hand side gives the familiar equal spacing rule of vector meson masses.
The second term now superimposes an adjoint BEH phenomenon over this equal 
spacing rule. For this mechanism to work, we must have the inequalities
$$
 \Delta_{ij}:= m_{ij}^2- (m_i+m_j)^2 = m_{ij}^2- \frac{(m_{ii}+m_{jj})^2}{4} 
\geq 0   ~~~~~~~~ i,j=1,...N
\eqno(4)
$$

The last expression before the inequality sign determines $\Delta_{ij}$ in 
terms of the experimentally known values of the vector meson masses, so that
many of these inequalities can be tested. Let us label the six quark 
flavors by their usual
symbols $u, d, s, c, b, t$ rather than by the numbers 1,2,...,6.
In view of 
$m_\rho^\pm = m_\rho^0 = m_\omega= 770-783$ MeV, it is not necessary to
distinguish the $u$ and $d$ flavors, so  we can use the common 
notation $n$ for 
both of these flavors. Using the experimentally known values
$m_\rho=0.77 ~{\rm GeV},~m_\omega=0.782 ~{\rm GeV},~
m_{\phi}=1.02 ~{\rm GeV},~m_{J/\psi}=3.097~{\rm GeV}, ~m_{\Upsilon}=9.46 ~{\rm GeV},~m_{{\rm K}^*}=0.892~{\rm GeV},~m_{{\rm D}^*}=2.01~{\rm GeV},~m_{{\rm D}_{s}^*}=2.112~{\rm GeV},~m_{{\rm B}^*}=5.325~{\rm GeV}$, we then find 
\setcounter{equation}{4}
\begin{eqnarray}
\Delta_{ns}=-0.005~{\rm GeV}^2,&~~~~~&\Delta_{nc}=0.30~{\rm GeV}^2,\nonumber \\
\Delta_{sc}=0.22~{\rm GeV}^2,&~~~~~&\Delta_{nb}=2.13~{\rm GeV}^2.
\end{eqnarray}
Though $\Delta_{ns}$ is negative, this is irrelevant given that 
$|\Delta_{ns}| < (m_{\omega})^2 - (m_{\rho})^2$, which has been neglected
as explained above.
In other words $\Delta_{ns}=0$ at the level of accuracy 
intended here. Since
$$\Delta_{ij}= \left( \frac{(\theta_i - \theta_j) R'}{2\pi \alpha'} \right)^2,
\eqno(6)
$$
it then follows that not only do we have $\theta_u = \theta_d$ but even the 
stronger equalities $\theta_u = \theta_d = \theta_s$ This in turn requires 
$\Delta_{nc} =\Delta_{sc}$, which holds to within 27\% or so.
The picture that emerges has three D-branes, the $u$, $d$ and $s$ branes 
essentially
coinciding, while the remaining three branes are further away from them.
Quantitatively
\setcounter{equation}{6}
\begin{eqnarray}
  |\theta_d - \theta_u| R' &=& |\theta_s - \theta_u| R'~~=~~0, \nonumber \\
  |\theta_c - \theta_u| R' = 3.44 ~{\rm{GeV}}^{-1}, ~~~&~&~~~
  |\theta_b - \theta_u| R' = 9.17~ {\rm{GeV}}^{-1}.
\end{eqnarray}
At the present stage of top spectroscopy
$|\theta_t - \theta_u| R'$ is not determined. 
The D-brane arrangement corresponding to these $\theta_i$'s 
is presented in figure 1. The mass formula (3) accounts well for the observed 
vector meson masses (there is an experimentally less constrained spin $0$ 
counterpart to all this).
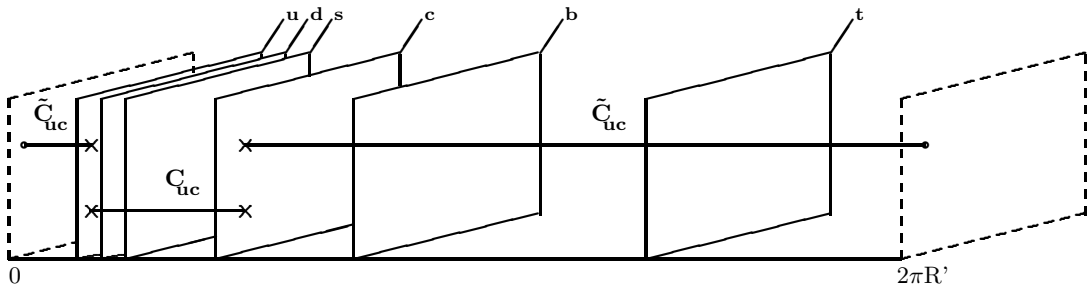
\begin{figure}[t]
\setlength{\unitlength}{0.011500in}%
\begingroup\makeatletter
% extract first six characters in \fmtname
\def\x#1#2#3#4#5#6#7\relax{\def\x{#1#2#3#4#5#6}}%
\expandafter\x\fmtname xxxxxx\relax \def\y{splain}%
\ifx\x\y   % LaTeX or SliTeX?
\gdef\SetFigFont#1#2#3{%
  \ifnum #1<17\tiny\else \ifnum #1<20\small\else
  \ifnum #1<24\normalsize\else \ifnum #1<29\large\else
  \ifnum #1<34\Large\else \ifnum #1<41\LARGE\else
     \huge\fi\fi\fi\fi\fi\fi
  \csname #3\endcsname}%
\else
\gdef\SetFigFont#1#2#3{\begingroup
  \count@#1\relax \ifnum 25<\count@\count@25\fi
  \def\x{\endgroup\@setsize\SetFigFont{#2pt}}%
  \expandafter\x
    \csname \romannumeral\the\count@ pt\expandafter\endcsname
    \csname @\romannumeral\the\count@ pt\endcsname
  \csname #3\endcsname}%
\fi
\endgroup
\begin{picture}(490,125)(32,500)
\thicklines
\multiput( 60,575)(-0.41667,-0.41667){13}{\makebox(0.4444,0.6667){\SetFigFont{7}{8.4}{rm}.}}
\multiput( 55,575)(0.41667,-0.41667){13}{\makebox(0.4444,0.6667){\SetFigFont{7}{8.4}{rm}.}}
\multiput(130,575)(-0.41667,-0.41667){13}{\makebox(0.4444,0.6667){\SetFigFont{7}{8.4}{rm}.}}
\multiput(125,575)(0.41667,-0.41667){13}{\makebox(0.4444,0.6667){\SetFigFont{7}{8.4}{rm}.}}
\multiput(130,605)(-0.41667,-0.41667){13}{\makebox(0.4444,0.6667){\SetFigFont{7}{8.4}{rm}.}}
\multiput(125,605)(0.41667,-0.41667){13}{\makebox(0.4444,0.6667){\SetFigFont{7}{8.4}{rm}.}}
\multiput( 60,605)(-0.41667,-0.41667){13}{\makebox(0.4444,0.6667){\SetFigFont{7}{8.4}{rm}.}}
\multiput( 55,605)(0.41667,-0.41667){13}{\makebox(0.4444,0.6667){\SetFigFont{7}{8.4}{rm}.}}
\put( 27,603){\circle{2}}
\put(27,603){\line(1,0){30.5}}
\put(437,603){\circle{2}}
\put(127,603){\line(1,0){310}}
\put(57,573){\line(1,0){70}}
\multiput(426,551)(0.00000,7.68421){10}{\line( 0, 1){  3.842}}
\multiput(426,624)(7.25,1.81){12}{\curve(0,0,4,1)}
\multiput(510,644)(0.00000,-7.68421){10}{\line( 0,-1){  3.842}}
\multiput(426,551)(7.25,1.81){12}{\curve(0,0,4,1)}
\put(424,540){\makebox(0,0)[lb]{\smash{\SetFigFont{9}{10.8}{rm}
2$\pi$R'
}}}
\put(310,551){\line( 4, 1){ 83.765}}
\put(310,551){\line( 0, 1){ 73}}
\put(310,624){\line( 4, 1){ 83.765}}
\put(394,644){\line( 0,-1){ 73}}
\put(393,573){\line( 0, 1){  0}}
\put(394,645){\line( 2, 3){ 10}}
\put( 31,614){\makebox(0,0)[lb]{\smash{\SetFigFont{9}{10.8}{bf}
\~{C}
}}}
\put( 36,610){\makebox(0,0)[lb]{\smash{\SetFigFont{7}{8.4}{bf}
uc
}}}
\put( 91,584){\makebox(0,0)[lb]{\smash{\SetFigFont{9}{10.8}{bf}
C
}}}
\put( 96,580){\makebox(0,0)[lb]{\smash{\SetFigFont{7}{8.4}{bf}
uc
}}}
\put(285,614){\makebox(0,0)[lb]{\smash{\SetFigFont{9}{10.8}{bf}
\~{C}
}}}
\put(290,610){\makebox(0,0)[lb]{\smash{\SetFigFont{7}{8.4}{bf}
uc
}}}
\put(293,551){\makebox(0.4444,0.6667){\SetFigFont{10}{12}{rm}.}}
\put( 20,551){\line( 1, 0){406}}
\put( 51,551){\line( 0, 1){  0}}
\multiput( 51,551)(0.55135,0.09189){21}{\makebox(0.4444,0.6667){\SetFigFont{7}{8.4}{rm}.}}
\put( 62,551){\line( 0, 1){  0}}
\multiput( 62,551)(0.55135,0.09189){21}{\makebox(0.4444,0.6667){\SetFigFont{7}{8.4}{rm}.}}
\put( 51,624){\line( 4, 1){ 83.765}}
\put(135,644){\line( 0,-1){  1}}
\put(262,577){\line( 0,-1){  6}}
\put( 62,551){\line( 0, 1){ 73}}
\put( 62,624){\line( 4, 1){ 83.765}}
\put(146,644){\line( 0,-1){  1}}
\put(177,551){\line(4,1){83.765}}
\put(177,551){\line( 0, 1){ 73}}
\put(177,624){\line( 4, 1){ 83.765}}
\put(262,644){\line( 0,-1){ 73}}
\put(262,573){\line( 0, 1){  0}}
\put( 51,624){\line( 0,-1){ 73}}
\multiput(20,551)(7.25,1.81){4}{\curve(0,0,4,1)}
\curve(49,558.24,51,558.74)
\multiput( 20,551)(0.00000,7.68421){10}{\line( 0, 1){  3.842}}
\multiput(20,624)(7.25,1.81){12}{\curve(0,0,4,1)}
\multiput(104,644)(0.00000,-5.33333){2}{\line( 0,-1){  2.667}}
\put(114,551){\line( 4, 1){ 62.824}}
\put(114,551){\line( 0, 1){ 73}}
\put(114,624){\line( 4, 1){ 83.765}}
\put(198,644){\line( 0,-1){ 14}}
\put(198,630){\line( 0, 1){  0}}
\put( 73,551){\line( 4, 1){ 41.176}}
\put( 73,551){\line( 0, 1){ 73}}
\put( 73,624){\line( 4, 1){ 83.765}}
\put(157,644){\line( 0,-1){ 10}}
\put(157,633){\line( 0, 1){  0}}
\put(135,645){\line( 2, 3){ 10}}
\put(146,645){\line( 2, 3){ 10}}
\put(157,645){\line( 2, 3){ 10}}
\put(198,645){\line( 2, 3){ 10}}
\put(262,645){\line( 2, 3){ 10}}
\put(278,551){\line( 1, 0){145}}
\put( 20,540){\makebox(0,0)[lb]{\smash{\SetFigFont{9}{10.8}{rm}
0
}}}
\put(146,660){\makebox(0,0)[lb]{\smash{\SetFigFont{7}{8.4}{bf}u}}} 
\put(157,660){\makebox(0,0)[lb]{\smash{\SetFigFont{7}{8.4}{bf}d}}}
\put(168,660){\makebox(0,0)[lb]{\smash{\SetFigFont{7}{8.4}{bf}s}}}
\put(209,660){\makebox(0,0)[lb]{\smash{\SetFigFont{7}{8.4}{bf}c}}}
\put(273,660){\makebox(0,0)[lb]{\smash{\SetFigFont{7}{8.4}{bf}b}}}
\put(405,660){\makebox(0,0)[lb]{\smash{\SetFigFont{7}{8.4}{bf}t}}}
\end{picture}
\vspace*{-1.5cm}
\caption{Configuration of flavor D-branes.}
\end{figure}

It may be worthwhile at this point to take notice of the remarkable changes of
Minkowski space under T-duality. Whereas in the ``original" string picture
with compactification radius $R$, there was a unique ambient extended 
Minkowski space in which all strings --- and therefore all string ends ---
moved, in the T-dual picture with compactification radius $R'$
we find $N$ Dirichlet 3-branes on which all
strings must end. It is as if Minkowski space in this picture were 
$N$-sheeted, the number $N$ of sheets being determined by the number 
of quark flavors. {\em A geometrical interpretation of the number of quark 
flavors 
thus emerges}. We should stress that this interpretation is tied to the 
hadronic string considered here and therefore to the large 
number of quark colors picture which leads to it. It is not to be confused
with similar space foliations in a fundamental string theory \cite{P}, 
where the 
basic scale is dictated by the much smaller Planck length. 
Multi-sheeted Minkowski space
is also encountered in noncommutative geometry based approaches to the 
Standard Model \cite{C}.

Though the mass formula (3) is quite realistic, our main interest in this 
picture comes from the fact that it accounts naturally 
for the universal coupling 
of vector mesons, usually referred to as vector meson dominance.
It has been known for a very long time indeed \cite{S} that the 
vector mesons considered here couple as if they were flavor $U(N)$ 
gauge bosons.
For instance the $\rho^0$ couples twice as strongly to $\pi^+$ mesons as to
protons, as expected for a gauge boson of isospin, 
the third component of the $\pi^+$-meson's isospin being double that of the 
proton. This coupling pattern has not been derived from QCD so far. 
The {\em true} gauge bosons of QCD are of course the gluons, and in QCD 
the vector 
mesons are composite states whose coupling pattern and mass spectrum
are to be worked out numerically, say by lattice QCD calculations. 
The string picture considered here 
has the advantage of automatically yielding the right coupling 
pattern for the vector mesons, which appear here as massive gauge bosons (in 
the limit of vanishing quark masses and coincident D-branes for all flavors,
they become massless and $U(N)$ gauge invariance is manifest). 
The string picture also corrects 
the unacceptable mass spectrum produced by a pure 
adjoint BEH phenomenon for these
vector mesons. The only other work which deals with both these 
vector meson problems is 
that based on hidden local symmetries of a nonlinear sigma model
\cite{BK}. It is
not clear whether or how this work relates to the string approach taken here.
\bigskip

{\bf 3. The Magnitude of the Compactification Radius and the Glueball Problem}

\bigskip

We have now come to the point where we have to confront the question whether
the fifth dimension introduced in these arguments is at all acceptable.
To meaningfully discuss this problem we first have
to estimate the compactification radius $R$ or its dual $R'$. This can be 
done in the  following manner. Beside the string stretched between the 
D-branes $i$ and $j$, which contributes the $\Delta_{ij}$ term to the 
square of the vector 
meson mass, there is another ``complementary" string which goes in the opposite
direction along the $\theta$ circle, e.g. the complementary string of the 
string marked $C_{uc}$ in 
figure 1 is the string ${\tilde{C}}_{uc}$.
 For this string not to correspond to a lighter state, we need
$2\pi - |\theta_i - \theta_j| \geq |\theta_i - \theta_j|$ or $|\theta_i - \theta_j| \leq \pi$ for all $i$ and $j$.
Therefore we must have
$$
    R' \geq \frac{1}{\pi}\max(|\theta_i - \theta_j| R')
\eqno(8)
$$
 From Eq. (5), using Eqs. (3) and (4), it can be seen that 
$|\theta_b - \theta_n| R'/|\theta_c - \theta_n|R'\approx 2.7$, which 
is very close to $m_b/m_c=m_{\Upsilon}/m_{J/\psi}=3.06$, so that it appears 
that $|\theta_i - \theta_j| R'$ increases more or less linearly
with the (absolute value of the) mass difference of the flavor $i$ and 
flavor $j$ quarks, at least up to the $b$ quark. We do not know how this 
extends to the top quark and will therefore consider two cases:
i) $|\theta_t - \theta_n| R'/|\theta_b - \theta_n|R'\approx \frac{m_t}{m_b}$ 
and ii) $|\theta_t - \theta_n| R'/|\theta_b - \theta_n|R'\approx O(1)$.

In case i) the maximum in Eq.(8) sets in for $i=t$ and $j=u$, 
and, using $m_t=175$ GeV, this maximum 
can be estimated at some $108$ GeV$^{-1}$, so 
$R' \geq 21$ fermi! If the spacing of the levels of the mesonic KK tower were 
determined by $R'^{-1}$, this would certainly be the end of the story. 
{\em For open strings}
the spacing of the levels of the mesonic 
KK tower is dictated \cite{P} by $R^{-1}$, with $R$ 
given by Eq. (1), i.e. $R \leq \frac{1}{108}$ GeV$^{-1} \approx 0.002$ fermi.
In other words, on the ``$R$ side" we are dealing with with a 
compactification scale comparable to the weak scale, 
{\em or smaller}. On the ``$R'$ side", on the other hand, the large scale
of 20 Fermi appears and one would expect this scale, or the equivalent  
$\frac{1}{R'}=10$ MeV mass scale,
also to manifest itself 
in the theory. It does indeed and in a disastrous way at the level of closed
strings, i.e. of glueballs. Such closed string glueballs 
are an inescapable consequence 
of unitarity (which forbids a pure open string theory). 
Unlike the open 
strings, these closed string glueballs can wind around the small radius $R$
circle and give winding modes. If the lightest glueball has mass $m_G$
(typically around 2 GeV), then these $R$-winding, or equivalently 
$R'$-KK modes involve 
the masses $\sim m_G(1+n^2/R'^2 m_G^2)^\frac{1}{2},~~~n=1,2,3,...$. 
These modes 
are very closely spaced, with the resulting glueball sector hardly 
distinguishable from a continuum. Though the couplings of these glueball
states to ordinary $q\overline{q}$ mesons are OZI suppressed, by their sheer
number these glueball states would 
dominate production processes to an unacceptable degree.
In case ii) the details change somewhat: $10$ MeV becomes $\sim 1$ GeV, but 
this glueball disaster persists.
It is in the glueball sector that the theory betrays its five-dimensional 
nature.
Though at the level of open strings everything worked out for the best
at least at experimentally available energies, at 
the glueball level the continuous fifth dimension cannot be tolerated.
This glueball-flood is due primarily to the fact that a continuous fifth 
dimension produces infinite KK towers, which cannot be cut off.
To cut off the KK towers, 
we next explore the possibility that the fifth dimension 
introduced in this stringy hadron phenomenology is really {\em discrete},
somewhat along the lines of noncommutative geometry \cite{C}.

\bigskip

{\bf 4. A Discrete Fifth Dimension}

\bigskip

As suggested at the end of the previous section, we now proceed to 
explore the consequences of a discrete fifth dimension. As we saw, the string 
theory has two T-dual pictures, involving circles of radii $R$ and $R'$. Here
we replace these two circles by two periodic lattices:
one with $N$ points, period $2\pi R$, 
and spacing $a$; the other with $N'$ points, period $2\pi R'$, 
and spacing $a'$. This way, 
$Na= 2\pi R$ and $N'a'=2\pi R'$. Because of the periodicity of the lattice, 
on the ``unprimed" side the fifth component of the momentum 
carried by closed strings is quantized in 
units of $\frac{1}{R}$. 
Thus the KK modes carry fifth component of the momentum 
$\frac{n}{R}=\frac{2\pi n}{Na}, ~n\in {\bf Z}$, 
whereas the winding modes carry fifth 
component of the momentum $\frac{wR}{\alpha'}=\frac{w}{R'}~, ~w\in {\bf Z}$. 
Because we are dealing with a
lattice, the possible values of the fifth component of the momentum 
are in the interval $[-\frac{\pi}{a},\frac{\pi}{a}]$. For the extremal values
in this interval to be compatible with the KK-quantization of the momentum, we 
must choose $N$ a positive even integer. The KK modes then 
range over the integer multiples of $\frac{1}{R}$ in the interval 
$[-\frac{N}{2R}, \frac{N}{2R}]$ and we also have
$|\frac{wR}{\alpha'}|\leq \frac{N}{R}$.
Similar statements, with primed and 
unprimed quantities interchanged, hold on the 
primed side as well. The KK (winding) modes on the unprimed side are to
be identified with the winding (KK) modes on the primed side with
$w\longleftrightarrow n'$ and $w'\longleftrightarrow n$.
Again for the extremal modes, this requires
$$
\frac{N}{R}=\frac{N'}{R'},
\eqno(9a)
$$
or equivalently 
$$
a=a'.
\eqno(9b).
$$
The spacings of the primed and unprimed lattices must be same.

The T-duality condition on the compactification radii, Eq. (1), now becomes
$$
\frac{4\pi^2 \alpha'}{a^2}=NN',
\eqno(10)
$$
so that the square of the lattice spacing is rationally related to $2\pi$
times the string tension.
When the fifth dimension is discrete, Eq. (6) becomes
$$
|\nu'_i-\nu'_j|a=2\pi \alpha' \sqrt{|\Delta_{ij}|},
\eqno(11)
$$
where $\nu'_i$ is the position of the flavor $i$ membrane 
on the primed lattice.
Eqs. (10) and (11) yield
$$
|\nu'_i-\nu'_j|=\sqrt{NN'\alpha'|\Delta_{ij}|}.
\eqno(12)
$$
Eqs. (10)-(12) will be very useful towards obtaining the 
lattice spacing and the sizes of the two lattices. 
First of all we get an idea of possible values of $N'$. Since all 
$|\nu'_i -\nu'_j|\leq \frac{N'}{2}$, Eqs. (10) and (12) require
$$
 N' \geq 4N\alpha'\max |\Delta_{ij}|,
\eqno(13)
$$
the discrete equivalent of Eq. (8).
Using the values of $\alpha'$ given above and of the largest $|\Delta_{ij}|$
known so far, namely $|\Delta_{bn}|$ (see Eq. (5)), 
we find $N'\geq 16N$. 
We are thus naturally led to the ``minimal" case $N=2$ and to values of $N'$ 
in excess of $16$.
As a first example,
set $N=2,~~ N'=50$ (both even as was explained above). Using 
$\alpha'\approx 1$ GeV$^{-2}$ we find
$a=1.26\times10^{-14}$ cm and with $|\nu'_b-\nu'_n|=15$ we obtain the value 
$|\Delta_{bn}|=2.25$ GeV$^2$, which closely reproduces the values in Eq. (5).
We will find other interesting lattice configurations in the next section.
Before we discuss these further choices of the lattice parameters, let us now
see how this discrete fifth dimension solves the glueball problem of the 
continuous case. In the continuous case the KK towers were infinite and given
their small spacing the production processes were swamped with glueballs.
In the lattice case, by contrast, the KK tower consists of $N'$ states, i.e.
of a {\em finite} number of states.
If the lowest glueball state's mass is $m_G \sim 2$ GeV, then its KK tower
will consist of $N'$ states of masses
$m_n =\sqrt{m_G^2 +(\frac{2\pi n}{N'a})^2}$ where the integer $n$ runs from
$-\frac{N'}{2}$ to $+\frac{N'}{2}$. For the just discussed example
this means 51 states in the mass interval $2$ - $5.4$ GeV. 
The production of each
of these states is OZI suppressed by say a factor 0.1. We would  therefore
expect the production of a continuum-like spectrum,
but not with the immense cross-section
obtained when the fifth dimension was a continuous circle.

We should mention that both for open and closed strings, besides the KK towers,
one expects the characteristic exponentially growing string-like spectrum. 
A string-like 
exponentially exploding hadron
spectrum is indicated by experiment, as was already known to Hagedorn 
\cite{H}.
The absence of tachyons places severe constraints on the Fermi-Bose
imbalance of this spectrum \cite{KS}, which again seem to be obeyed 
\cite{FR, CD} in the region
in which hadron spectroscopy is reliable.
\bigskip

{\bf 5. Tests of the Hadronic String Model with a Discrete 
 Fifth Dimension}

\bigskip

We saw in section 3 the constraints which follow in the continuous case 
from the requirement that the mesons correspond to the lighter of a
``complementary" pair (e.g. strings $C_{uc}$ and $\tilde{C}_{uc}$ in fig.1).
These complementary pairs exist in the discrete case as well. For the $[ij]$ 
meson which extends over $|\nu'_i-\nu'_j|$ primed 
lattice sites, its complementary string extends 
over $N'-|\nu'_i-\nu'_j|$ sites. 
Though heavier, this complementary $\tilde{C}_{ij}$ string should nevertheless 
exist. Its mass is given by
$$
\tilde{m}_{ij}^2= (m_i+m_j)^2 +\left(\frac{(N'-|\nu'_i-\nu'_j|) a}{2 \pi \alpha'}\right)^2
\eqno(14)
$$
We also record here the discrete analog of Eq. (3) for the mass of the 
``original" (shorter) meson states:
$$
m_{ij}^2= (m_i+m_j)^2 +\left(\frac{|\nu'_i-\nu'_j| a}{2 \pi \alpha'}\right)^2
\eqno(15)
$$
One can see from the last two equations that,
as in the continuous case, both the ``original" and the ``complementary" string
carry fractional amounts of the unprimed KK quantum of the fifth component
of the momentum. Such a fractional charge is possible in the familiar fashion
\cite{P} because of the $U(N)$ and of the D-branes it entails.
Note that for ``diagonal"
$[ii]$ type mesons, like the $\rho,~ \Upsilon, $ etc., Eq. (14) yields 
their first (and for $N=2$ their last) {\em unprimed} KK mode. 
We now have to see how the existence of these complementary 
strings is reflected in meson spectroscopy. We must compare their predicted 
masses with meson spectroscopic data. We proceed in four steps:

--- a) We set $N=2$ and choose for $N'$ an even integer $\geq 16$; we also 
impose $N'\leq 70$ so as not to obtain an overly crowded glueball spectrum, 
in line with our discussion in Section 4 above.

--- b) From Eq. (10) we determine the lattice spacing $a$.

--- c) From Eqs. (15) and (4) we determine the best integers 
$|\nu'_i-\nu'_j|\leq \frac{N'}{2}$ which fit the observed $\Delta_{ij}$ of 
Eq. (5).

--- d) We insert the results of the previous three steps into Eq. (14) and obtain the complementary meson masses.

The results of the first three steps are displayed in Table 1 and the 
complementary meson masses 
obtained in step d) are
in Table 2. 
\begin{table}[h]
\centering
\begin{tabular}{|c|c|c|c|}\hline
 $N'$ & a & $|\nu'_c-\nu'_n|$ & $|\nu'_b-\nu'_n|$\\
& (fermi)& &\\ \hline
20 & 0.199 & 3 & 9 \\
30 & 0.162 & 4 & 11 \\
40 & 0.140 & 5 & 13 \\
50 & 0.126 & 5 & 15 \\
60 & 0.115 & 6 & 16 \\
70 & 0.106 & 6 & 17 \\ \hline
\end{tabular}
\caption{Lattice Parameters}
\end{table}
\begin{table}[ht]
\centering
\begin{tabular}{|c|r|r|r|r|}\hline
 $N'$ & ${\tilde{m}}_{J/\Psi}$ & ${\tilde{m}}_{\Upsilon}$ & $\tilde{m}_D$ & $\tilde{m}_B$ \\ 
     & (GeV) & (GeV) & (GeV) & (GeV)\\ \hline 
  20 & 3.62 & 9.97 & 3.31 & 5.40\\
 30 & 4.25 & 10.22 & 3.87 & 5.67 \\
 40 & 4.81 & 10.46 & 4.36 & 5.94 \\
 50 & 5.30 & 10.70 & 4.90 & 6.20 \\
 60 & 5.75 & 10.93 & 5.30 & 6.50 \\
 70 & 6.17 & 11.16 & 5.74 & 6.80 \\ \hline
\end{tabular}
\caption{Complementary Meson Masses}
\end{table}

As far as the mesons complementary to
the light mesons are concerned, these 
are all predicted to have masses between $3$
and $6$ GeV, where no meaningful spectroscopic data are available. 
Therefore in Table 2
we show only the heavy mesons. For these  we obtain masses which are not 
much above the heaviest observed mesons and for the $b\bar{b}$ case 
are actually in the 
accessible range. The typical decay modes of these complementary mesons is to
the ``original mesons" (whose complementaries they are) and a real or virtual 
glueball which then decays to pions. These complementary mesons are not
to be confused with the radial excitations of the ``originals". As can be 
seen from Table 2, the most promising sectors where complementary strings 
might be found are those of the $B$, $B_s$ and $b\bar{b}$ mesons.

We again stress that the glueball proliferation is solved in all the cases of 
Tables 1 and 2 in the same way as outlined in the last section.

By inspecting Tables 1 and 2 we can observe some important features.
First of all, the ratio $\frac{|\nu'_b-\nu'_n|}{|\nu'_c-\nu'_n|}$ is close to $3$ for 
all choices of $N'$, as  was already noted in the continuous case in Section
3. Next we notice that the values obtained for $\tilde{m}_{\Upsilon}$  tend to
favor the larger values of $N'$, those above $40$. This raises a number of 
questions. First of all, if the distance between the $t$ and $uds$ D-branes
were much larger than that between the $b$ and $uds$ D-branes, then much 
larger $N'$ values would come into play ($N'\geq 300$). With such large $N'$
the glueball proliferation problem would resurface.
With the values of $N'$ 
in the tables we are constrained to the alternative ii) of Section 3. 

All this
raises the much deeper question as to how the values of $N'$ and of the 
$\nu'_i$,
the locations of the D-branes on the primed lattice, are determined. We have 
no insight at present into this important question. The comparable question 
for the unprimed lattice has been disposed of by making the ``minimal" choice
$N=2$. $N'$ between 20 and 70 is then somewhat removed from the 
self-dual point. For fundamental strings this would be hard to understand, 
but then we are {\em not} dealing with fundamental strings here.

\bigskip

{\bf 6. Conclusions}

\bigskip

In this paper we proposed a string picture of hadrons, as suggested by the
large number of colors limit of QCD. This has automatically brought 
with it the Chan-Paton rules which are so useful in understanding
the phenomenology of high energy hadron scattering at fixed momentum
transfer and therefore of hadronic total cross-sections. Expanding Minkowski
space by an additional discrete space dimension, we were led to a simple
picture of vector meson masses and couplings. This picture led us to
the prediction of ``complementary" mesonic strings which would be heavier than
the usual mesons and would decay into these ordinary mesons and glueballs.
In the $\Upsilon$, $B$ and $B_s$ sectors such complementary mesons lie
sufficiently close to the experimentally accessible region as to make
a search for such states possible. 

To get agreement with the meson spectrum, we were led to a set of values for
the parameters which describe the discrete fifth dimension: a lattice 
spacing of $\sim 10^{-14}$ cm and periodic lattices with 2 points in one
picture and 20-70 points in the picture T-dual to it. 
It would be very interesting to
understand the theoretical reasons for the appearance of these particular 
lattices.

We should stress that on account of the discrete fifth dimension, 
we dealt throughout with
a four dimensional situation with an enriched spectrum. There are {\em 
cut off} KK towers but all that does not produce the kind of proliferation
encountered in the presence of a continuous fifth dimension where the KK 
towers would become infinite.

\bigskip

{\bf Acknowledgments}
\bigskip

We wish to thank Emil Martinec, for valuable discussions.

This paper was supported in part by NSF grant PHY-9123780-A3.


\newpage
\begin{thebibliography}{99}

\bibitem{RD} C. Vafa, hep-th/9702201.

\bibitem{DR} J. Schwarz, hep-th/9607201.

\bibitem{GSW} M. B. Green, J.H. Schwarz and E. Witten, 
{\em Superstring Theory}, vol. I, Cambridge University Press, Cambridge 1987.

\bibitem{PF} P.G.O. Freund, Phys. Rev. Lett. {\bf 20}, 235 (1968).

\bibitem{HH} H. Harari, Phys. Rev. Lett. {\bf 20}, 1395 (1968).

\bibitem{HHR} H. Harari, Phys. Rev. Lett. {\bf 22}, 562 (1969).

\bibitem{JR} J.L. Rosner, Phys. Rev. Lett. {\bf 22}, 689 (1969).

\bibitem{FFP} J.A. Feigenbaum, P.G.O. Freund and M. Pigli, 
preprint EFI-97-13, hep-ph/9703296

\bibitem{FR} P.G.O. Freund and J.L. Rosner, Phys Rev. Lett. 
{\bf68}, 765 (1992).

\bibitem{TH} G. 't Hooft, Nucl. Phys. {\bf B72}, 461 (1974)

\bibitem{EW} E. Witten, Nucl. Phys. {\bf B160}, 571 (1979)

\bibitem{P} J Polchinski, hep-th/9611050, to be published.

\bibitem{S} J.J. Sakurai, Ann. Phys. NY {\bf 11}, 1 (1960).

\bibitem{CT} A. Chodos and C, Thorn, Nucl. Phys. {\bf B72}, 509 (1974). 

\bibitem{C} A. Connes, {\em Noncommutative Geometry}, Academic Press, NY, 1994.

\bibitem{CC} A. Chamseddine and A. Connes, hep-th/9606001.

\bibitem{BK} M. Bando, T. Kugo, S. Uehara, K. Yamawaki and T. Yanagida,
Phys. Rev. Lett. {\bf 54}, 1215 (1985).

\bibitem{H} R. Hagedorn, Nuovo Cimento {\bf56A}, 1027 (1968).

\bibitem{KS} D. Kutasov and N. Seiberg, Nucl. Phys. {\bf B358}, 600 (1991).

\bibitem{CD} J.R. Cudell and K.R. Dienes, Phys. Rev. Lett. 
{\bf 69}, 1324 (1992).







\end{thebibliography}
\end{document}